\documentclass[%
 reprint,
 amsmath,amssymb,
 aps,
]{revtex4-1}

\usepackage{graphicx}
\usepackage{dcolumn}
\usepackage{bm}
\newcommand{\overbar}[1]{\mkern0.9mu\overline{\mkern-0.9mu#1\mkern-0.9mu}\mkern 0.9mu}
\usepackage{color}
\begin{document}

\preprint{AIP/123-QED}

\title[PAPER SCHEME]{Laser induced fluorescence for axion dark matter detection: \\ a feasibility study in YLiF$_4$:Er$^{3+}$}

\author{C. Braggio}
 \email{caterina.braggio@unipd.it}
\author{G. Carugno}%
\author{F. Chiossi}
\email{federico.chiossi@phd.unipd.it}
\affiliation{ 
Dip. di Fisica e Astronomia and INFN, Sez di Padova, Via F. Marzolo 8, I-35131 Padova, Italy
}%

\author{A. Di Lieto}
\affiliation{Dip. di Fisica and INFN, Largo Bruno Pontecorvo, 3, I-56127 Pisa, Italy}

\author{M. Guarise}
\affiliation{ 
Dip. di Fisica e Astronomia and INFN, Sez di Padova, Via F. Marzolo 8, I-35131 Padova, Italy
}%

\author{P. Maddaloni}
\affiliation{CNR-INO, Istituto Nazionale di Ottica, Via Campi Flegrei 34, I-80078 Pozzuoli, Italy}
\affiliation{INFN, Istituto Nazionale di Fisica Nucleare, Sez. di Napoli, Complesso Universitario di M.S. Angelo, Via Cintia, 80126 Napoli, Italy}

\author{A. Ortolan}
\author{G. Ruoso}
\affiliation{INFN, Laboratori Nazionali di Legnaro, Viale dell'Universit\`a 2, I-35020 Legnaro, Italy}

\author{L. Santamaria}
\affiliation{Agenzia Spaziale Italiana (ASI), Contrada Terlecchia, I-75100 Matera, Italy}

\author{J. Tasseva}
\affiliation{INFN, Istituto Nazionale di Fisica Nucleare, Sez. di Napoli, Complesso Universitario di M.S. Angelo, Via Cintia, 80126 Napoli, Italy}

\author{M. Tonelli}
\affiliation{Dip. di Fisica and INFN, Largo Bruno Pontecorvo, 3, I-56127 Pisa, Italy}

\date{\today}

\begin{abstract}

We present a detection scheme to search for QCD axion dark matter, that is based on a direct interaction between axions and electrons explicitly predicted by DFSZ axion models. The local axion dark matter field shall drive transitions between Zeeman-split atomic levels separated by the axion rest mass energy $m_a c^2$. 
Axion-related excitations are then detected with an upconversion scheme involving a pump laser that converts the absorbed axion energy ($\sim $ hundreds of $\mu$eV) to visible or infrared photons, where single photon detection is an established technique. 
The proposed scheme involves rare-earth ions doped into solid-state crystalline materials, and the optical transitions take place between energy levels of $4f^N$ electron configuration. 
Beyond discussing theoretical aspects and requirements to achieve a cosmologically relevant sensitivity, especially in terms of spectroscopic material properties, we experimentally investigate backgrounds due to the pump laser at temperatures in the range $1.9-4.2$\,K. Our results rule out excitation of the upper Zeeman component of the ground state by laser-related heating effects, and are of some help in optimizing activated material parameters to suppress the multiphonon-assisted Stokes fluorescence.

Valid PACS numbers may be entered using the \verb+\pacs{#1}+ command.
\end{abstract}

\pacs{Valid PACS appear here}
\maketitle

\section{\label{sec:level1} Introduction}
The nature of particle dark matter (DM) is the most long standing question in Big Bang cosmology, and direct searches may shed light on this intriguing mystery. The non-detection of DM in the heavy mass range (10\,GeV to 10\,TeV) \cite{Aprile:2017,Akerib:2017,Agnese:2017} has motivated the scientific community to focus theoretical and experimental efforts on much lower mass particles \cite{Essig:2012zr,Kouvaris:2017vn}. 
Among them, a well motivated light particle is the QCD axion \cite{Wilczek:1978uq,Weinberg:1978vn}, introduced by Peccei-Quinn to solve the strong CP problem \cite{Peccei:1977ly}. 
Axion physical properties are described by several models that can be grouped into the KSVZ \cite{Kim:1979kx,Shifman:1980} and DFSZ  classes \cite{Zhitnitsky:1980,Dine:1981,Dine:1983uq}, depending on zero or full axion coupling strength to leptons, respectively.    Even so, an almost model-independent statement holds for the axion mass  
\begin{equation}
\label{axmass}
m_a \simeq 0.6\times 10^{-4}\,{\rm eV} \left(\frac{10^{11} {\rm GeV }}{f_a} \right), 
\end{equation} 
 where  $f_a$ is  the Peccei-Queen symmetry-breaking energy scale, inversely proportional to the coupling strenghts with standard model particles  \cite{Wilczek:1978uq,Weinberg:1978vn}.  A light and stable axion emerges as an ideal DM candidate if  large $f_a$ are considered.  Due to the resulting huge occupation number, galactic halo axions can be described as a classical oscillating field $a$, with oscillation frequency $\nu_a=m_a c^2/h$ \cite{Sikivie:1983kl}.
The $10^{-6}<m_a<10^{-3}$\,eV axion mass range has since long been favoured by astrophysical and cosmological bounds \cite{raffelt:2008}, while very recent high-temperature lattice QCD calculations suggest that $m_a\geqslant 50\,\mu$eV \cite{Borsanyi:2016fk}.

The axion is intensively searched in \emph{haloscope} experiments \cite{Marsh:2016ve}, mostly based on resonant axion-photon conversion in a static magnetic field via Primakoff effect \cite{Sikivie:1983kl}. 
The  Axion Dark Matter eXperiment (ADMX) is the most sensitive haloscope detector based on  high quality factor  microwave resonators at cryogenic temperature.   
ADMX searches have excluded the mass range $1.9<m_a<3.69\,\mu$eV \cite{Asztalos:2010kx,Sloan:2016uq}. The experiment HAYSTAC (formally ADMX-High Frequency)   \cite{Brubaker:2017qf}, designed specifically to search for axions in the $20-100\,\mu$eV range ($5-25$\,GHz), has recently reached cosmologically relevant sensitivity at 24\,$\mu$eV ($\sim$5.8\,GHz). 

The  axion-electron coupling, explicitly predicted by DFSZ models \cite{Zhitnitsky:1980,Dine:1981,Dine:1983uq}, can be considered  to envisage another class of haloscopes, thereby providing the opportunity to discriminate among axion models in case of detection. Complementary approaches may prove crucial to determine the fractional amount of axions as DM constituent.  
 For instance, inhomogeneous filled cavities, in which the effective axion field is converted to magnetization oscillations of a ferrimagnet, are under study \cite{Barbieri:2017dq}. 
In this case, single photon detection is required, and it can be realized by e.g. superconducting circuit devices acting as quantum bits properly coupled to the cavity photons \cite{Wallraff:2004nx, Tabuchi405}, but as yet their dark count rate still exceeds the axion interaction rate. 

Approaches described so far are affected by an extremely poor sensitivity for axion masses above 0.2 meV ($\sim 50$ GHz),  where the effective detector volume is a critical issue.  Extension to the mass range up to 1 meV (250 GHz) may be rather accomplished in suitable condensed matter experiments, in which the space parameters hardly accessible to cavity technology could be tackled with the upconversion scheme investigated in this work, whereby cosmological axions cause transitions between Zeeman split levels of suitable atomic species.  

As target atoms we consider rare-earth (RE) elements inserted as dopants in crystalline matrices, where they exist as trivalent ions, substitutional for one of the atoms of the host with the same valence state and similar ionic radius. 
Among RE ions, those with an odd number of $4f$ electrons are called Kramers ions \cite{spectro:2005}, and have electronic doublet levels with magnetic moments of the order of $1-10$ Bohr magnetons $\mu_B$. Therefore, using Kramers doublets, axion-induced spin transitions can take place in the GHz range with application of moderate magnetic fields.  
  For instance, in Er$^{3+}$, the calculated splitting spans from 20 to 120\,GHz with applied magnetic fields in the interval 0.4 to 2.5\,T \cite{Marino:2016}, which translates to a large tunability in the favoured cosmological axion mass window. 

\begin{figure*}[t]
\includegraphics[width=5.5in]{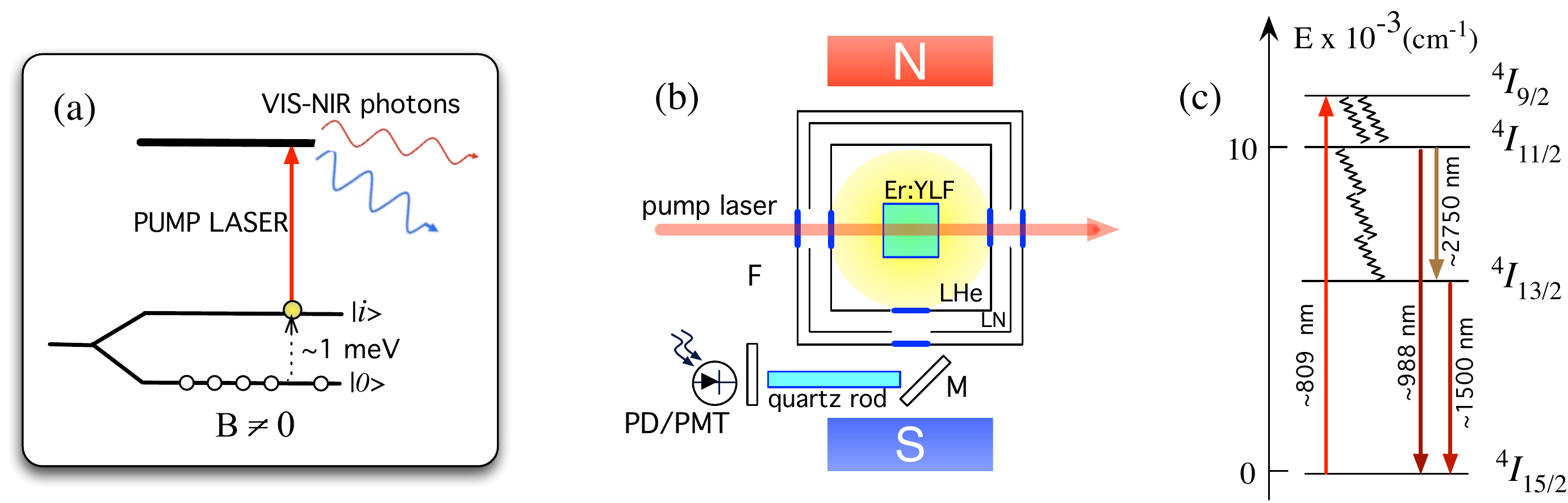}
\caption{\label{fig:1}
(a) Detection scheme: axion induced transitions take place between the Zeeman split ground state levels, then a laser pumps the excited atoms to a fluorescent level.
(b) Laser-induced fluorescence experimental setup. During the tests the crystal is immersed LHe and superfluid He. Fluorescence is collected orthogonally to the laser propagation direction by means of a mirror (M) that couples light to a 10\,cm-long quartz guide. Optical filters (F) are set in front of the InGaAs photodiode (PD) or photomultiplier tube (PMT) to remove stray light.
 (c) Portion of the energy level diagram of YLiF$_4$:Er$^{3+}$ and transitions that are relevant for the present work \cite{Tkachuk:2002uq}. Downward arrows indicate fluorescence transitions when ground state absorption takes place at about 809\,nm laser pump wavelength.}
\end{figure*}

In the direct axion-electron coupling 
\cite{Krauss:1985uq,Barbieri:1989vn} the interaction energy is $ (g_{ae}/2e)\vec{\nabla}a \cdot \vec{\mu}$, where the term $(g_{ae}/2e)\vec{\nabla}a$ plays the role of an effective oscillating field, $\vec{\mu}$ is the electron magnetic moment with electric charge $e$ and $g_{ae}$ is the coupling constant \cite{Marsh:2016ve}.  
Resonant condition is met when the Zeeman splitting energy is $m_ac^2$. 
  As schematized in Fig.\,\ref{fig:1}\,(a), the axion excitation is upconverted by a pump laser to photons in visible or infrared ranges, where single photon detection with ultra-low dark count rate has been already demonstrated \cite{Hadfield:2009kx,Schuck:2013fk,Dreyling-Eschweiler:2015cr}. 
The proposed detection scheme is based on electronic transitions between states within a $4f$ configuration of the trivalent RE, with positions of the discrete energy levels minimally perturbed by the crystal-field due to the screening action of the $5s$ and $5p$ orbitals \cite{spectro:2005}. 
It is immediately evident that a first requirement for the feasibility of such a scheme is related to the 
the linewidth of the transition driven by the laser, which must be narrower than the energy difference between the atomic levels $ | 0 \rangle$ and $ | i \rangle$. 

Detectability of axions in this scheme can be at first discussed by considering only the thermal excitation of the atomic level as fundamental noise limit. Backgrounds of different nature are left for experimental investigations in the second part of the work.
We consider one mole of target atoms in the ground state $ | 0 \rangle$ and, using Eq.\,8 of Ref.\,[\onlinecite{Sikivie:2014fk}], we establish the transition rate to the level $ | i \rangle$ by axion absorption on resonance:
   \begin{equation}
   \begin{aligned}
\label{rate}
N_AR_i= & \, 8.5\times 10^{-3} \, \left( \frac{\rho_a}{0.4\,{\rm GeV/cm}^3}   \right) 
\left( \frac{E_a}{330\,\mu{\rm eV}}   \right)^2 \\
& \cdot g_i^2 
 \left(\frac{\overbar{v^2}}{10^{-6}}  \right) \left(\frac{{\rm min}(t,\tau,t_{\nabla a})}{10^{-6}\,\rm s}  \right)\,\rm Hz,
\end{aligned}
\end{equation}
where $R_i$ is the transition rate of a single target atom, $N_A$ is the Avogadro number, $E_a=h\nu_a$ is the axion energy,
$g_i$ is the coupling strength to the target atom and is of the order of one \cite{Sikivie:2014fk}, and $\overbar{v^2}$ is the mean square of the axion velocity.  
The value $330\,\mu$eV (80\,GHz) is a midpoint of the Zeeman splitting frequency interval reported for Er$^{3+}$ in Ref.\,\onlinecite{Marino:2016}.
As the in the considered galactic halo model axions are the dominant component of dark matter, we take for its energy density $\rho_a$ the value  0.4\,GeV/cm$^3$ obtained from the rotational curves.

    The experiment coherence time is set by ${\rm min}(t,\tau,\tau_{\nabla a})$, where $t$ is the measurement integration time (inverse of the resolution bandwidth), and $t_{\nabla a}$ is the axion gradient coherence time at the resonant frequency of the experiment that can be calculated  from the axion coherence time 
    $\tau_a =h/(E_a  \overbar{v^2}/c^2)$ \cite{Turner:1990}. The latter is related to the width of the axion kinetic energy distribution in the laboratory frame. If  we assume a  Maxwellian velocity distribution  in the Galactic rest frame and we take  $ (\overline{v^2})^{1/2} \simeq 10^{-3} c$ as the local dark matter virial velocity,  we get $\tau_a = 91 \cdot (330\,\mu {\rm eV}/m_a)\,\mu {\rm s}$, and finally \cite{Barbieri:2017dq}
 \begin{equation}
\label{tau}
\tau_{\nabla a}\simeq 0.68 \, \tau_a =   4 \left(\frac{330\,\mu{\rm eV}}{E_a}   \right)\left( \frac{Q_a}{1.9\times 10^6}  \right) \mu{\rm s} . 
\end{equation} 
where the merit factor $Q_a \equiv 2 \tau_a \nu_a \simeq 1.9\times 10^6$ qualifies the axion-microwave line width in haloscope experiments. 
 
The lifetime of the Zeeman excited state $\tau$ is typically much longer than $\tau_{\nabla a}$, and in the rare-earth doped materials considered in this work is strongly dependent on temperature, intensity of the static magnetic field, dopant concentration \cite{Macfarlane:1987,Thiel:2011uq, Lutz:2016dq,Cruzeiro:2017qf}. The magnetic field, beyond splitting degenerate levels and thus opening a channel for resonant axion detection, may also inhibit spin flips and thus increase the lifetime $\tau$ of the intermediate level. Lifetimes much longer than ms have been measured in several rare-earth activated optical materials at LHe temperature with magnetic fields comparable to those used in this work ($\sim 0.5$\,T) up to about 3\,T \cite{Thiel:2011uq}. 
Incidentally, for a given pump laser intensity, the efficiency of the mentioned upconversion process is greater for longer $\tau$, thus allowing for mitigation of the laser power requirements when large detecting volumes are devised \cite{Krupke:1965}. 

   As one might expect, the experiment must be operated in a ultra-cryogenic environment to minimize thermal population of the Zeeman excited level. To establish the working temperature of the apparatus, we treat the pumped crystal as if it were a single photon detector with overall efficiency $\eta=0.5$ (including the efficiency of  upconversion, the fluorescence collection efficiency and self absorption), and calculate the allowed thermal rate $R_t^{\prime}$ by requiring that the signal to noise ratio (SNR) is at least 3.  
    This condition is equivalent to $R_t^{\prime}\sim 4.8 \times 10^{-3}$\,Hz for an observation time of 1 hour \cite{Lamoreaux:2013ys}. For a given temperature $T$ of the doped crystal, the thermal excitation rate is related to the lifetime of the Zeeman excited level 
    \begin{equation}
R_t=\bar{n}/\tau,
\end{equation}  
with $\bar{n}=(1+\exp{(E_a/kT}))^{-1}$ average number of excited ions in the energy level $E_a$, and $k$ the Boltzmann constant.  
It is worth noticing that the contribution of adjacent Stark sublevels (due to interaction with the crystalline field) is not considered when their energy is much higher than $E_a$, as the case analyzed in this work. If a level lifetime of $\tau=1$\,ms is taken, we eventually get $\bar{n}\leqslant 5\cdot 10^{-6}$ for the allowed thermal rate $R_t^{\prime}$ and thus 80\,GHz mass axions can be searched provided the active crystal is cooled down to at least 300\,mK. 

 In the following we present a systematic investigation of possible backgrounds induced by the pump laser in Er$^{3+}$-doped YLiF$_4$ crystals at cryogenic temperatures and sub-Tesla magnetic field.  

 \section{\label{exp} Experimental}
 
 The $5\times 5 \times5$\,mm$^3$ volume, Er$^{3+}$:YLiF$_4$ crystals used in this work were grown with the Czochralski method. They have nominal Er$^{3+}$-dopant concentration of 1\,\% and 0.01\% concentration (atomic percent substitution for Y$^{3+}$). These concentrations correspond to $1.4\cdot10^{20}$ and $1.4\cdot10^{18}$ ions/cm$^{3}$, with three $4f$ electrons for each ion available as axion targets. 
Crystals with low concentration of dopants have been the subject of much scientific investigation for photon-echo-based optical data storage and data processing, owing to their 
narrow $4f-4f$ transition linewidths and long optical coherence times (see \cite{Thiel:2011uq} and references therein). In this work we are interested in the behavior of higher concentration samples to maximize the axion interaction rate given by Eq.\,(\ref{rate}) for a given laser-pumped, active detector volume. Moreover, the 1\% concentration samples allow for higher sensitivity to laser related backgrounds in the measurements described in sections \ref{noise} and \ref{multiphonon}.

  To allow for Zeeman studies at LHe and superfluid He temperatures, the samples were housed in an immersion dewar located between two NdFeB magnetic discs that produced a field of 370\,mT at the sample position. The c-axis of the crystal was parallel to the magnetic field direction. As shown in Fig.\,\ref{fig:1}\,(b), the sample fluorescence is collected orthogonally to the laser pump propagation direction and coupled to the photon detector by means of a mirror M and a quartz guide.  With optical filters we suppress scattered pump radiation at signal wavelengths, and at the InGaAs photodiode we detect the 1.5\,$\mu$m component of the overall infrared fluorescence spectrum (see Fig.\,\ref{fig:1}\,(c)). 
  The employed optical source is a cw Ti:sapphire laser, which can be finely tuned by rotating intracavity ethalons. Zeeman studies are conducted around 809\,nm wavelength (section \ref{GS}), while laser-induced backgrounds are investigated at 810.1\,nm (section \ref{noise}). The laser linewidth is $\delta \lambda \leqslant 2$\,GHz ($\sim 1-2$\,pm), comparable with the detected transitions widths. The incident light polarization angle is varied by means of a half-wave plate.
A typical value of laser intensity used in our measurements is 10\,W/cm$^2$, compatible with 0.1 upconversion efficiency in trivalent ions \cite{Krupke:1965}.
  For the laser noise studies described in section \ref{noise} the pump laser was chopped at 15 Hz to allow phase-sensitive detection. 
  
\section{\label{GS} Zeeman splitting of the $^{4}I_{15/2,\,5/2}$}

We consider the electronic ground state $^{2S+1}L_J$ of the erbium Kramers ion, the $^{4}I_{15/2}$. The interaction with the crystal-field splits the $2J+1$ magnetic sublevels into eight ($J+1/2$) Kramers doublets, labeled by $|M_J|$, namely the absolute value of the $J$ projection on the crystal optical axis $c$, $|M_J|=15/2, 13/2, \ldots , 3/2, 1/2$. Through application of a magnetic field parallel to the c-axis, each Kramers doublet splits into two magnetic components: $^{4}I_{15/2,|M_J|,-}$ and $^{4}I_{15/2,|M_J|,+}$. In YLiF$_4$:Er$^{3+}$, the lowest Kramers doublet is the  $^{4}I_{15/2,5/2}$. 
To determine its splitting under application of a 370\,mT magnetic field, we measure the wavelength of transitions coupling the Zeeman components of the ground $^{4}I_{15/2,5/2}$ and the $^{4}I_{9/2,\,9/2}$ excited level by laser induced fluorescence (LIF) measurements.
 \begin{figure}[h!]
\includegraphics[width=3.4in]{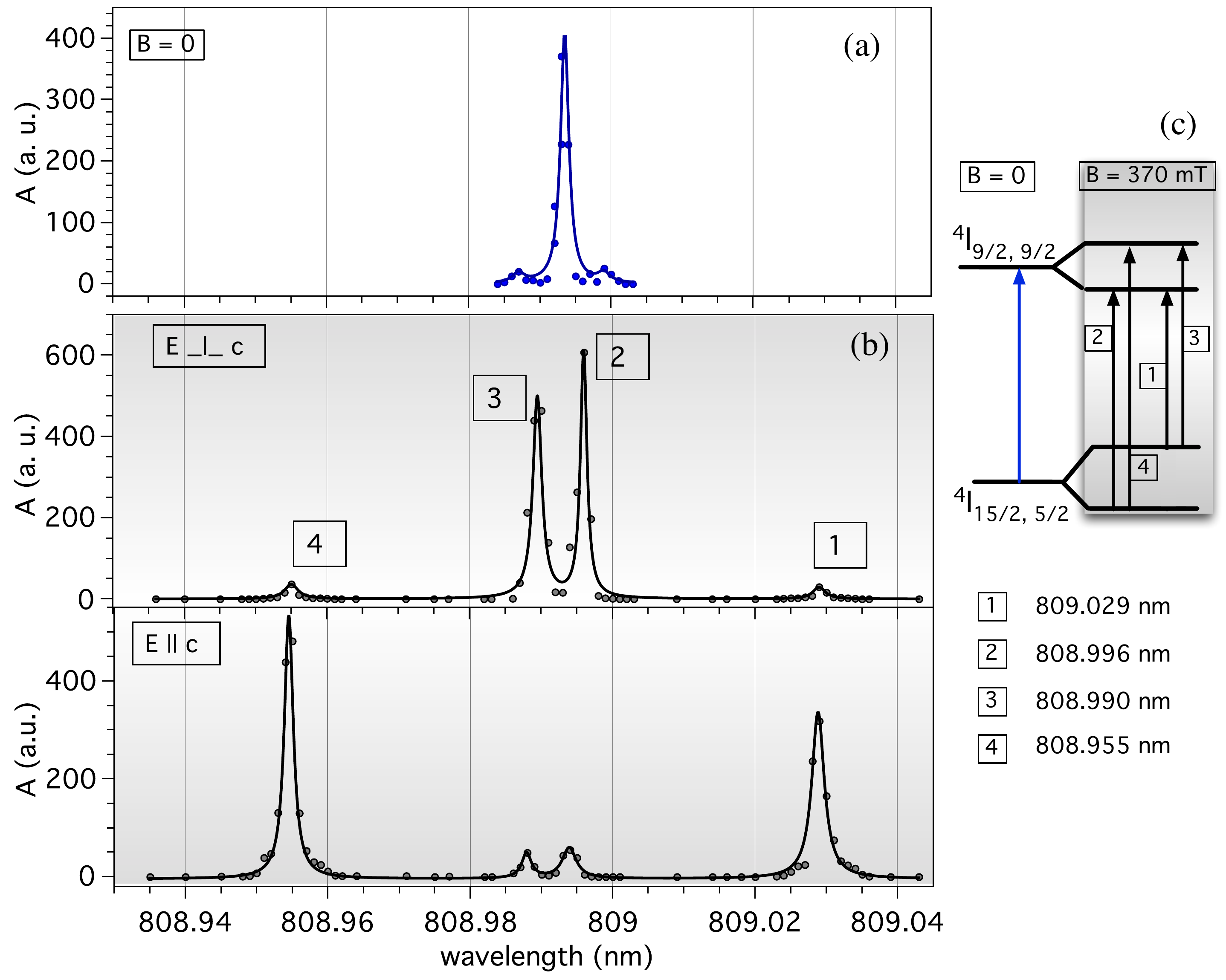}
\caption{\label{fig:2} 
The ground level ($^{4}I_{15/2,\,5/2}$) splitting with a magnetic field of 370\,mT is measured via laser induced fluorescence. During the measurements the 0.01\% concentration sample is immersed in liquid He. We report the registered fluorescence with no magnetic field (a) and with 370\,mT field (b) for laser polarization vector orthogonal or parallel to the crystallographic axis $c$.  The black line in these plots is only a guide to the eye.
Transitions between the Zeeman split levels of the ground state $^{4}I_{15/2,\,5/2}$ and the excited $^{4}I_{9/2,\,9/2}$ are identified as shown in part (c). }
\end{figure}
In Fig.\,\ref{fig:2} we report the results obtained with the 0.01\,\% concentration sample, where different incident laser polarization orientations (orthogonal and parallel to the crystallographic axis $c$) allow for a better detectability of possible transitions between Zeeman-split ground and excited states.
The fluorescence spectrum displays sharp, well-separated lines. Among the observed transitions, 
only $^{4}I_{15/2,\,5/2, +}\rightarrow\,^{4}I_{9/2,\,9/2, -}$   ($\lambda_1=809.029$\,nm) and $^{4}I_{15/2,\,5/2, -}\rightarrow\,^{4}I_{9/2,\,9/2, +}$ ($\lambda_4=808.955$\,nm) are unambiguously identified.  
With the displayed data it is then not possible to determine whether the energy levels differences  
$\Delta E_{31}=73.9$\,$\mu$eV, \,$\Delta E_{42}=77.7$\,$\mu$eV and 
$\Delta E_{21}=62.5$\,$\mu$eV,  \,$\Delta E_{43}=66.4$\,$\mu$eV 
represent the ground state or the excited level $^{4}I_{9/2,\,9/2}$ splittings. 
As described in section \ref{noise}, we accomplish this task by laser excitation of the thermal population in the Zeeman-split first excited Stark level of the ground state.  
Independently of this limitation, the plots in Fig.\,\ref{fig:2} demonstrate that we are able to resolve the 
Zeeman splitting and therefore that it is possible to 
monitor the population of the upper Zeeman component of the ground state. 
 Clearly at $T=2$\,K thermal excitation of the level still prevents us from assigning a detection sensitivity to the present apparatus, but before we get to cool the sample to hundreds of mK temperatures, a thorough investigation of the pump laser related noise is accomplished as described in the following sections. 
 
The LIF measurements in Fig.\,\ref{fig:2} have been repeated with the 1\,\% concentration sample. In this case the Zeeman transition is hardly resolved due to increased transition linewidths, ascribable to spin cross relaxation processes due to direct interactions among Er$^{3+}$ ions \cite{Bottger:2006fk,Cruzeiro:2017qf}. 
However, such a limitation might be overcome in the high magnetic field and low temperature regime, required to achieve ultimate sensitivity in the proposed axion detection scheme. 
For instance, in a 0.1\% concentration sample of YLiF$_4$:Er$^{3+}$, authors have investigated the four transitions connecting the Zeeman sublevels of the ground and lowest $^{4}F_{9/2}$ excited state and demonstrated that their linewidth can be as low as $\sim 1\,$MHz  \cite{Wannemacher:1989}. The applied magnetic field was about 3\,T and measurements were conducted below 4\,K  by Zeeman-switched optical-free-induction decay technique. 
These results, together with our findings, foster the development of a few liters detector with intermediate concentration active materials, matching the axion-induced transition rate in Eq.\,\ref{rate} to dark count rates in available single photon counters \cite{Hadfield:2009kx,Schuck:2013fk,Dreyling-Eschweiler:2015cr}. 
As a final additional remark, we note that an intermediate concentration sample would allow for increasing the axion-electron interactions of six orders of magnitude compared to a gaseous target prepared by buffer cooling techniques \cite{Santamaria:2015}.

\section{Laser-induced thermal noise}\label{noise}

To assess heating effects in the active detector volume, we focus on the population of the first excited Stark (crystal-field) sublevel $^{4}I_{15/2, 15/2}$, that has a strong thermal coupling with the ground energy level. To enhance the sensitivity of our tests, we use the 1\% concentration sample.
The crystal-field splittings of Er$^{3+}$ ions in YLiF$_{4}$ have been calculated and measured by previous authors \cite{Coutodos:1998,Kulpa:1975} and for the $^{4}I_{15/2, 15/2}$ sublevel the separation from the ground state is $E_{\rm{S}1}=17$\,cm$^{-1}$ (2.1077\,meV). If $E_{\rm{S}j}$ are the Stark sublevels energies of the $^{4}I_{15/2}$ level, the occupation probability of the first excited Stark level is proportional to $e^{-E_{\rm{S}1}/kT}/(1+\sum_{j=1}^{M_J} e^{-E_{\rm{S}j}/kT})$, where the sum is well approximated with the first two terms. 
As the fluorescence intensity $F$ is proportional to the laser intensity $I$ times the level occupation number $\bar n$, we model a possible heating effect with the term  $\beta I$ in expression $F(T,I) \simeq \alpha  I  \exp [-E_{\rm{S}1}/(kT+\beta I)]$, where $\alpha$, $\beta$ are empirical parameters determined from a fit to the data. 

\begin{figure}[h!]
\includegraphics[width=3.7in]{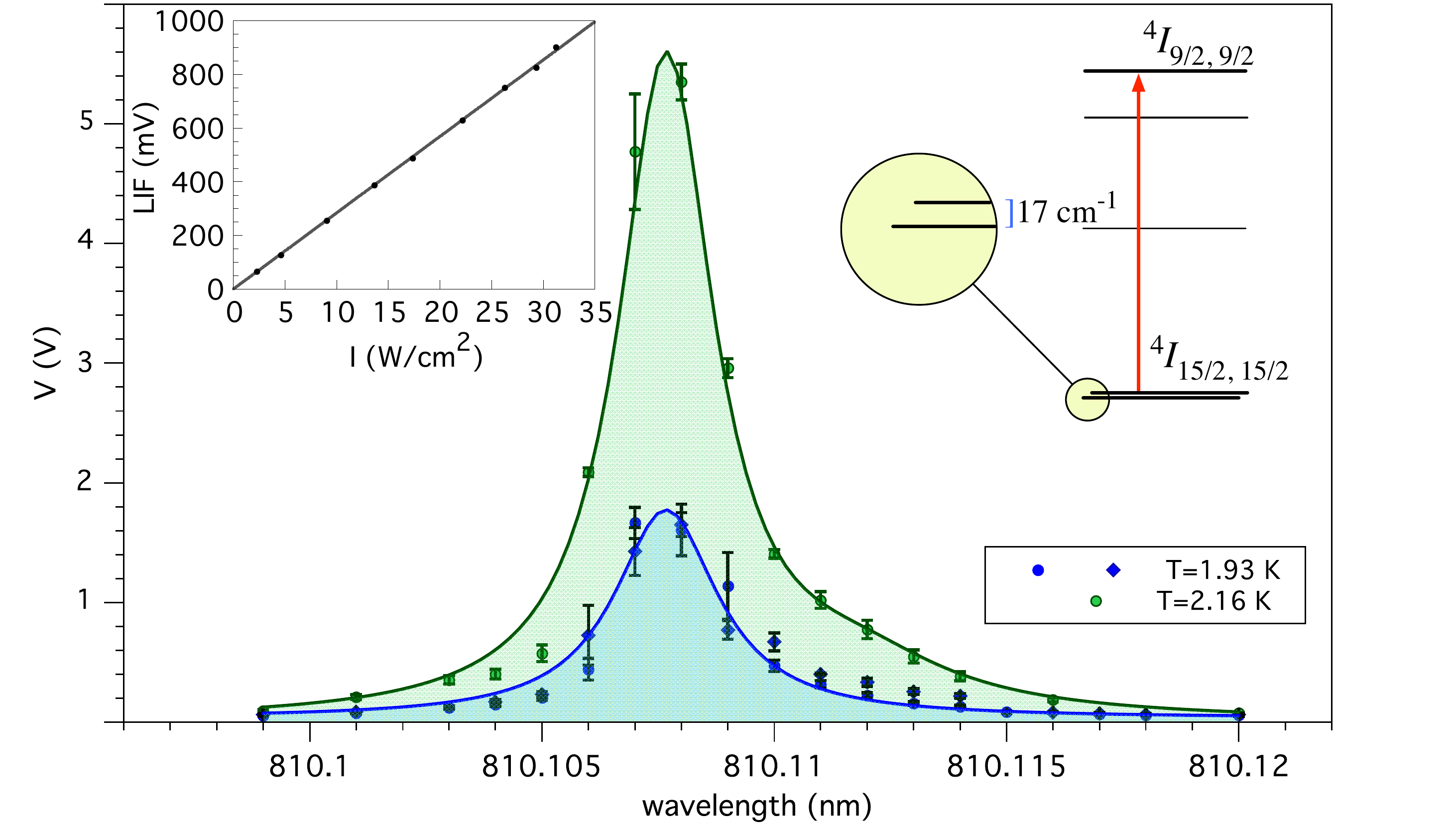} 
\caption{\label{fig:5} Probing the $^4I_{15/2,15/2}$ first excited Stark sublevel at 1.93 and 2.16\,K in the 1\,\% concentration sample. 
Inset shows a plot of the fluorescence amplitude for several pump laser intensity values measured at $T=4.2$\,K, with wavelength set at $\lambda=810.108$\,nm in resonance with the same Stark sublevel.  The linear fit rules out temperature increments greater than 0.2\,K/[W/cm$^2$] at $4\,K$, and allows in first approximation to assign the bath temperature to the laser-pumped crystal volume. The ratio between the two resonant peak areas is compared to the ratio of the Boltzmann factors calculated at 1.93 and 2.16\,K. 
}
\end{figure}

From the data shown in the inset of Fig.\,\ref{fig:5}, we infer that the $\beta$ parameter 
is compatible with zero within one standard deviation, which allows us to limit the temperature increase to less than 0.2\,K/[W/cm$^2$]. 
We stress that such limit is obtained in an unfavorable upconversion scheme, where the de-excitation takes place also through non radiative channels as shown in Fig.\,\ref{fig:1}\,(c). 
Therefore we can assign the temperature of the thermal bath to the entire crystal and calculate the ratio of the populations of the same Stark level at two different temperatures $T_1$ and $T_2$. Such ratio is then compared to the LIF peak areas. 
As shown in Fig.\,\ref{fig:5}, with the pump laser tuned to the transition $^{4}I_{15/2, 15/2}\rightarrow ^{4}I_{9/2, 9/2}$, we obtain peaks that differ only in their area parameter for $T_1=2.16$\,K and $T_2=1.93$\,K. 
The Er(1\%):YLF crystal is immersed in superfluid He, and these points are obtained under $\lambda$-point operation at which bubbling disturbances are eliminated. 
A small satellite line is present on the right side of the main peak at both temperatures, which hinders an accurate fitting of the data. Therefore we compare the areas of the main peaks at $T_1=2.16$\,K and $T_2=1.93$\,K by summing the amplitudes of the data recorded at four wavelengths around resonance.
 The ratio of $3.6\pm0.3$ is in agreement with the expected value, confirming the assumption made in the introduction to calculate the rate of excited atoms via thermal bath temperature. 

\begin{figure}[h!]
\includegraphics[width=3.4in]{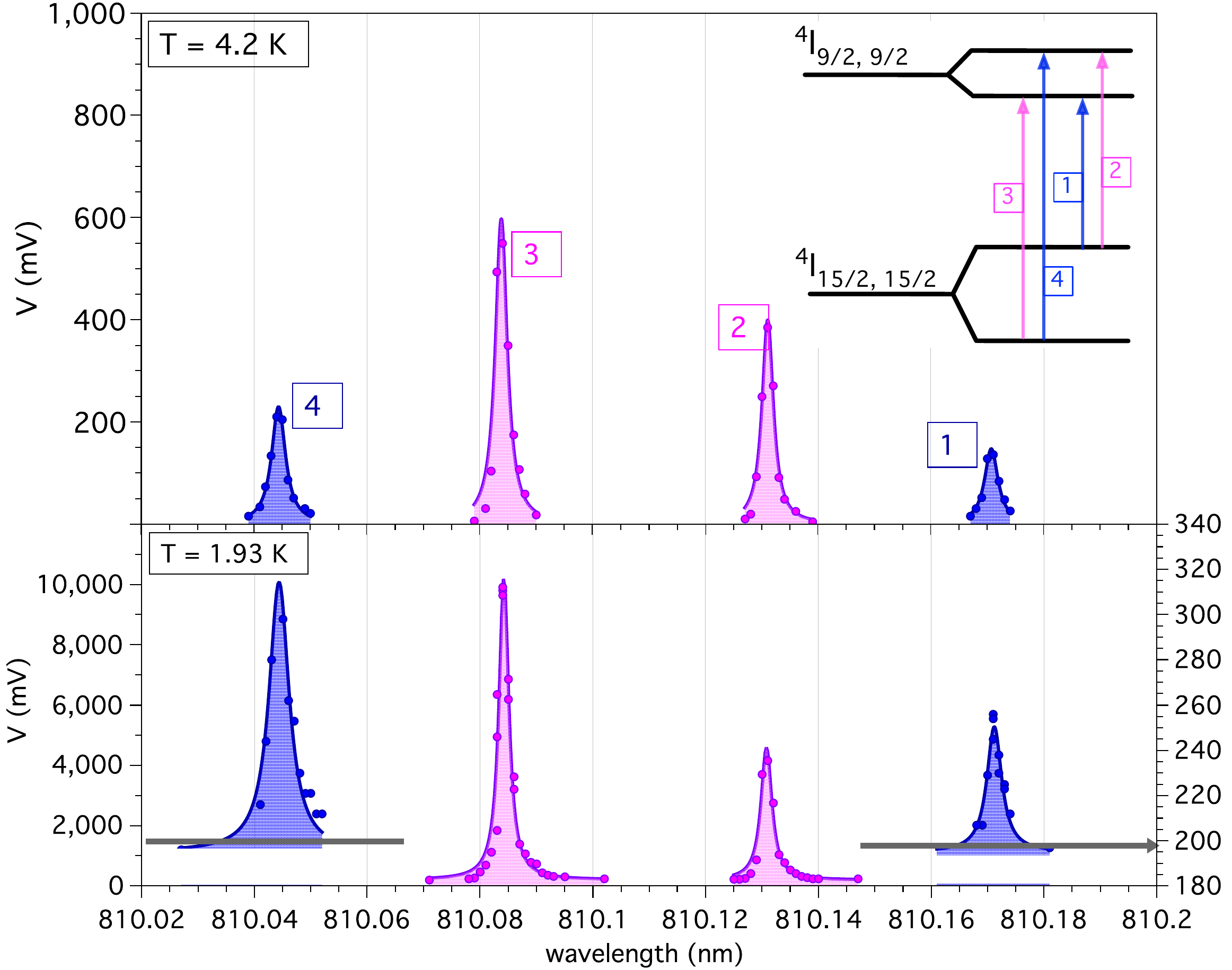}
\caption{\label{fig:4} Zeeman transitions from the first excited Stark level of the ground state $^{4}I_{15/2,\,15/2}$ to the lowest level of the excited $^{4}I_{9/2}$. The applied magnetic field is 370\,mT. Data sets corresponding to lines 1 and 4 at $T=4.2\,$K have been acquired with a different linear amplifier gain and lock-in amplifier sensitivity, and have to be divided by a factor 30 for direct comparison with lines 2 and 3. The physical origin of the measured background level ($\sim 195$\,mV) evidenced by the horizontal lines in the plot at $T=1.93\,$K is clarified in section \ref{multiphonon}. }
\end{figure}

 In addition, we confirm experimentally (Fig.\,\ref{fig:4}) that also the population of the Zeeman-split levels follows   Boltzmann statistics. 
 In this case, the pump laser is set to probe the populations of the Zeeman-split levels of the first excited Stark sublevel ($^{4}I_{15/2,\,15/2,+}$ and $^{4}I_{15/2,\,15/2,-}$).  As a larger splitting is expected for this level as compared to the ground state \cite{Kulpa:1975}, by pumping the transitions to the Zeeman levels of the $^4I_{9/2,9/2}$, the previously measured energy differences (see section \ref{GS}) can also be precisely identified.  In fact, from the wavelengths reported in the first column of Table\,\ref{tab:table1},  we obtain the splitting of the first excited Stark level
 $\Delta E_{e} \equiv \Delta E_{13}=\Delta E_{42}^{\prime}=164.4$\,$\mu$eV and $\Delta E_{12}^{\prime}=\Delta E_{43}^{\prime}=75.6$\,$\mu$eV, where the indices are assigned as described in the inset of Fig.\,\ref{fig:4}. The latter value is consistent with the a\-ve\-ra\-ge of $\Delta E_{31}=73.9$\,$\mu$eV, \,$\Delta E_{42}=77.7$\,$\mu$eV measured in section \ref{GS}. 
  Consequently, we take the ave\-rage $\Delta E_{g} \equiv (\Delta E_{21}+\Delta E_{43})/2=64.5\,\mu$eV as the searched splitting of the ground state. 
  To further confirm proper identification of the Zeeman split levels, we can use the ratios of reported $g$ factors in the same material oriented with its c-axis parallel to the magnetic field \cite{Kulpa:1975}. We obtain $\Delta E_{g}/\Delta E_{e}=0.39$, in agreement with $g_{||}[15/2,5/2]/g_{||}[15/2,15/2]=3.137/7.97=0.39\pm 0.01$.
    The ground state splitting value we measured at 0.37\,mT is also in fair agreement with theoretical values reported in Ref.\,\onlinecite{Marino:2016}. 
\begin{table}\small
\caption{\label{tab:table1} Lorentian fit of the data reported in Fig.\,\ref{fig:4}. The parameters $\lambda_c,\, A_i,\, \omega$ (center, area and width respectively) 
are expressed in nm, in (nm$\cdot$mV) and pm, respectively. 
 Errors on the peak areas are assigned by considering the error on the measured background at $T=1.93\,$K.}
\begin{ruledtabular}
\begin{tabular}{ccc|cc}
 &\multicolumn{2}{c}{$T=1.93$\,K}&\multicolumn{2}{c}{$T=4.2$\,K}\\
 $\lambda_c$ 
& $A_i$ & $w$  & $A_i$ 
& $w$ \\ \hline
1.\,\,810.171 & $0.25\pm0.03$ & $2.7\pm0.2$ & $0.671\pm 0.03$ & $2.9\pm 0.2$ \\
2.\,\,810.131& $16\pm0.2$ & $2.3\pm0.2$ & $1.6\pm0.1 $ & $2.4\pm 0.2$ \\
3.\,\,810.084 &  $32.8\pm$2 &$2.1\pm0.2$&$2.37\pm 0.23$ &$2.5\pm 0.2$\\
4.\,\,810.044 &$0.76\pm0.05$& $3.8\pm0.4$ & $1.05\pm0.03$   &$3.0\pm 0.2$\\
\end{tabular}
\end{ruledtabular}
\end{table}

That is as far as our LIF measurement of the ground level splitting is concerned. As for the investigations of possible laser-induced deviations from Boltzmann statistics, we consider 
the peak areas of LIF measured for the levels $^{4}I_{15/2,\,15/2,+}$ and $^{4}I_{15/2,\,15/2,-}$. 
The data displayed in Fig.\,\ref{fig:4} are fitted to a Lorentzian curve in the form $F(\lambda)=y_0+\displaystyle 2\frac{A_i}{\pi}\cdot \frac{w}{4(\lambda-\lambda_c)^2+w^2}$ and the results are 
reported in Table \ref{tab:table1}. The ratio $A_{1}/  A_{4}=0.69$ and  $A_{2}/ A_{3}=0.64$ is in fair agreement with 0.69 expected ratio at $T=4.2\,$K. At $T=1.93\,$K we obtain $A_{1}/  A_{4}=0.35$ and  $A_{2}/  A_{3}=0.48$, which averages to a value compatible with the calculated one 0.37.

\section{\label{multiphonon} Multiphonon background}
\begin{figure}[h!]
\includegraphics[width=3.1in]{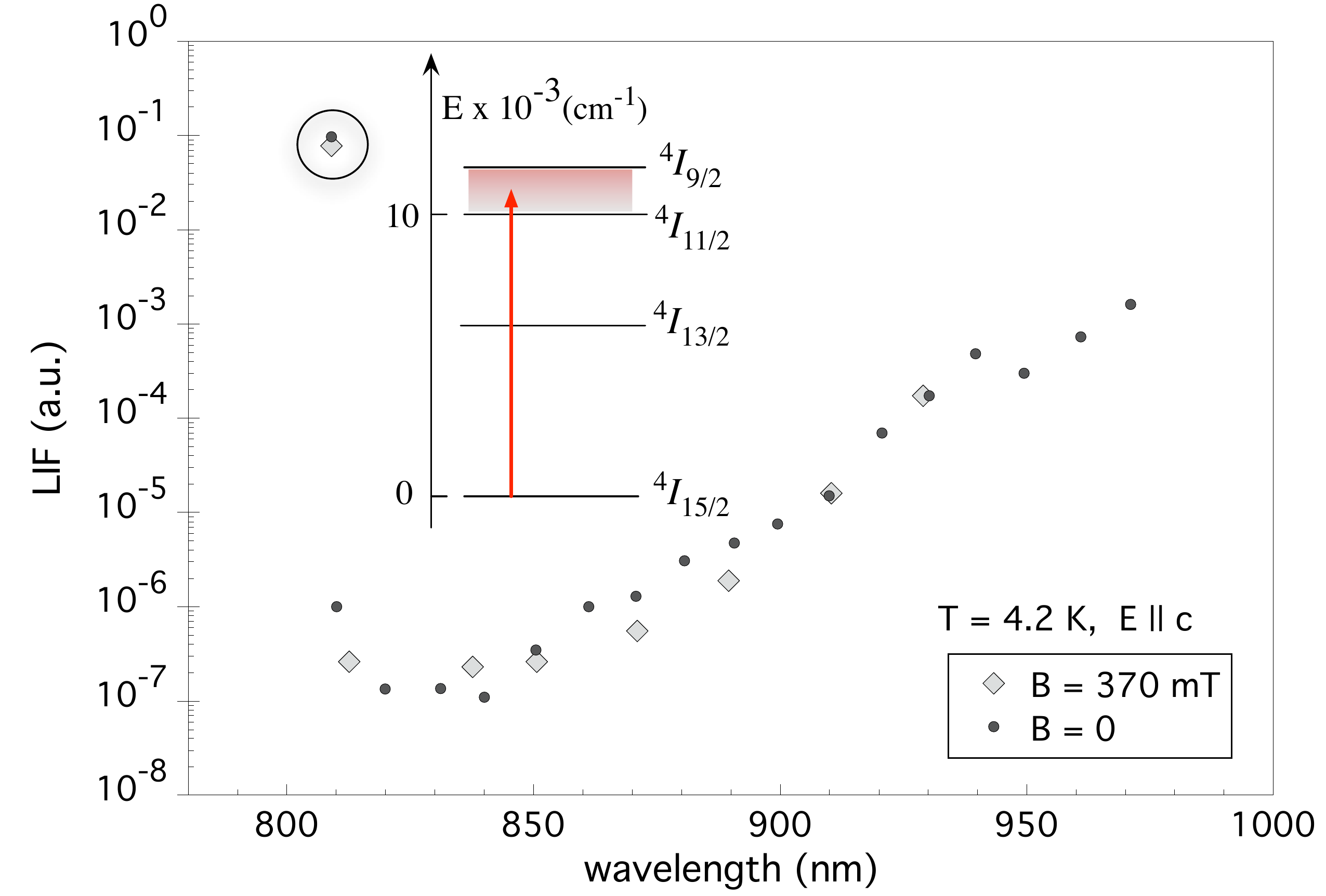}
\caption{\label{fig:Azuel} Observed fluorescence in the 10300--12360\,cm$^{-1}$ interval (shadowed band in the inset).  The LIF amplitude measured for pump laser resonant with transition $^{4}I_{15/2}\rightarrow^{4}I_{9/2} $ is shown for comparison (circled data). 
The out-of-resonance fluorescence is attributed to multiphonon-assisted anti-Stokes and Stokes emission. While the first component is evidently suppressed at $T=4\,$K, the Stokes fluorescence is represented in the plot by the exponentially increasing data.  
}
\end{figure}

In the data reported in Fig.\,\ref{fig:4} (lower side) the apparatus sensitivity allows observation of an out-of-resonance fluorescence level. To understand the physical origin of this background, we measure its intensity for values of pump laser wavelength in a wide range (corresponding to 10300--12360\,cm$^{-1}$), as shown in Fig.\,\ref{fig:Azuel}. The LIF measured with the pump laser 
tuned to transition $^{4}I_{15/2}\rightarrow^{4}I_{9/2}$ is also plotted for comparison.

A similar exponential behaviour has been previously reported in Er:YLF and has been explained in terms of multiphonon-assisted, side-band absorption \cite{Auzel:2004}. The RE manifolds $E_{1,2}$ can be in fact excited even by a nonresonant pump photon $E_1<E<E_2$, when the missing/excess energy is bridged by absorption/emission of phonons via Anti-Stokes and Stokes processes, respectively. 
The related absorbed intensity is theoretically given by:
\begin{equation}
\label{abs}
I(E)=I(E_1)e^{-\alpha_{\rm S}(E-E_1)}+I(E_2)e^{-\alpha_{\rm AS}(E_2-E)}
\end{equation}
where $\alpha_{\rm S}$ and $\alpha_{\rm AS}$ are the Stokes and Anti-Stokes coefficients, described in the model \cite{Auzel:1976} through expressions: 
\begin{eqnarray}
 \alpha_{\rm S} & =& (\hbar \omega_{\rm eff})^{-1}\ln \left (\frac{p}{S_0(n+1)}  -1  \right)  \label{1}\\
\alpha_{\rm AS} & = & \alpha_{\rm S}+1/kT.\label{2}
\end{eqnarray}

In Eq.\,(\ref{1}), $\hbar \omega_{\rm eff}$ is the crystal effective phonon energy, $p$ is the number of photons needed to bridge the energy gap, $n$ is the average occupation number
and $S_0$ the Huang-Rhys coefficient that represents the electron-phonon coupling strength.
Typical values of $\hbar \omega_{\rm eff}$ are smaller than $200$\,cm$^{-1}$ in bromides, greater than $400$\,cm$^{-1}$ in oxides \cite{Walsh:2016}, and in YLF $400$\,cm$^{-1}$ is reported.

 The rapid suppression of the LIF observed in our 4.2\,K data (Fig.\,\ref{fig:Azuel}) is ascribable to the expected suppression of the AS component with temperature (from eq.\,\ref{2}) and the exponential growth for increasing wavelengths is then mainly due to the Stokes process.  
 Fitting of data with wavelength greater than 850\,nm give an absorption coefficient $\alpha_{\rm S}=9.2\cdot 10^{-3}$\,cm$^{-1}$, in agreement with the value reported in Ref.\,\onlinecite{Chen:1995}.
Ground state absorption measurements allow to estimate a $2\cdot 10^{-20}$\,cm$^2$ cross section of the pure electronic transition (circled data in Fig.\,\ref{fig:Azuel}) and thus to quantify the upconversion efficiency and the multiphonon side band relative amplitude. This type of background hinders the application of the present scheme to axion detection, unless a suitable combination of rare-earth dopant, pumping pathway and matrix is chosen. In particular, relevant suppression of the background should be accomplished in low phonon energy host matrices \cite{Walsh:2016} or, as suggested by Eq.\,(\ref{abs}), by exploiting pumping schemes with larger $E-E_1$.    
It is worth mentioning that a ultimate laser-induced background might also originate from impurity absorption, the same process that is currently limiting the efficiency of optical refrigeration \cite{Seletskiy:2010,Melgaard:2014,Di-Lieto:2014}. 

\section{Discussion and conclusions}

We have discussed a solid-state approach for direct detection of axion dark matter, and established the most important experimental parameters necessary to reach cosmologically relevant sensitivity in DFSZ models. 
The effect of the continuous, coherent axion field is searched in the excitation of the Zeeman upper component  $ | i \rangle$ of the ground state of rare earth ions in crystalline matrices, at the energy scale $g_g\mu_{B} B$, corresponding to transitions in the $\sim 100\,$GHz range. 
 The population of this excited level is probed by a pump laser tuned to the transition to a fluorescent level within the same $4f$ atomic configuration, so as to convert the axion excitation into photons, detectable with state-of-the-art single-photon detectors.
 Assuming thermal excitation of the excited Zeeman level as fundamental noise limit, the active detector volume must be cooled down to ultracryogenic temperatures ($\lesssim 0.2$\,K). The rate of thermal excitation of the $ | i \rangle$ atomic level is directly related to its lifetime $\tau$, and the  temperature at which the final experiment must be performed has been estimated for 80\,GHz axion-induced transitions and $\tau=1$\,ms by requiring SNR$\gtrsim3$. 
As long as $\tau \geqslant 100\,\mu$s, upconversion with unitary efficiency is also ensured for 10\,W/cm$^2$ pumping intensity. 

 In the proposed scheme it is important to address a thorough experimental study of pump laser-related backgrounds. As a first step we have probed the population of atomic levels close to the ground state via LIF measurements in the temperature range $1.9-4.2$\,K. Our main finding is that the pump laser does not affect the thermal population of the Zeeman excited level at least up to a few W/cm$^2$  intensity. 
 In addition, we have shown that it is crucial to optimize the pumping pathway and crystal properties to minimize scattering of the pump photons on crystal phonons (Stokes process). 
 
  As for the detection scheme via laser induced fluorescence, at $4.2$\,K and with 370\,mT magnetic field, we have demonstrated that  the four transitions coupling the Zeeman levels of the ground and the excited $^{4}I_{9/2}$ can be resolved in the lowest concentration YLiF$_4$:Er$^{3+}$ sample (0.01\,\%). This was not possible in the 1\,\% sample. However, the spin population dynamics in Kramers ions strongly depends on applied magnetic field, temperature, dopant concentration and species, and we argue that a tradeoff between these parameters can be found for the proposed experiment feasibility. 
A few liters active volume ensures $\sim$\,mHz axion and thermal transition rates, corresponding to  statistically relevant counts of upconverted photons in a measurement time of a few hours. Fortuitously, the dark count rate of state-of-the-art single photon detectors holds below the transition rate in the detector active volume. 

 We are witnessing a blooming of table-top experiments pursuing new observables for axion DM  direct detection \cite{Budker:2014kx,Arvanitaki:2014zr,Rybka:2015bh,Kahn:2016ys,Silva:2017,Group:2017ve,CRESCINI2017109}. In such a multifaceted, dynamic scenario, our complementary proposal
  aims to probe the uncovered few hundred $\mu$eV axion mass region by exploiting the axion-electron interaction predicted in the DFSZ models.

\begin{acknowledgments}
The authors wish to thank I. Grassini for samples preparation. Technical support by E. Berto and F. Calaon is gratefully acknowledged.
\end{acknowledgments}

\nocite{*}

\providecommand{\noopsort}[1]{}\providecommand{\singleletter}[1]{#1}%
%


\end{document}